\documentclass[aps,prl,superscriptaddress,showpacs,floatfix]{revtex4-1}
\usepackage{graphicx,amsmath,amssymb,xspace,epsfig,float,multirow,subfigure,tabularx}
\usepackage{amsbsy,amssymb,amsmath,bm}
\usepackage{epsf}
\usepackage{color,natbib}
\usepackage{amsfonts}
\usepackage{hyperref}

\pdfoutput=1

\begin{document}

\title{Superconductivity in type II layered Weyl semi-metals }
\author{B. Rosenstein}
\affiliation{Department of Electrohysics, National Yang Ming Chiao Tung University, Hsinchu,
Taiwan, R.O.C. }
\author{B. Ya. Shapiro }
\affiliation{Department of Physics, Institute of Superconductivity, Bar-Ilan
University, 52900 Ramat-Gan, Israel.}

\begin{abstract}
Novel "quasi two dimensional" typically layered (semi) metals offer a unique
opportunity to control the density and even the topology of the electronic
matter. In intercalated $MoTe_{2}$ type II Weyl semi - metal the tilt of the
dispersion relation cones is so large that topologically of the Fermi
surface is distinct from a more conventional type I. Superconductivity
observed recently in this compound [Zhang et al, 2D Materials\textbf{\ 9,}
045027 (2022)] demonstrated two puzzling phenomena: the gate voltage has no
impact on critical temperature, $T_{c},$ in wide range of density, while it
is very sensitive to the inter - layer distance. The phonon theory of
pairing in a layered Weyl material including the effects of Coulomb
repulsion is constructed and explains the above two features in $MoTe_{2}$.
\ The first feature turns out to be a general one for any type II
topological material, while the second reflects properties of the
intercalated materials affecting the Coulomb screening.
\end{abstract}

\maketitle

\section{Introduction.}

\textit{\ }The 3D and 2D topological quantum materials, such as topological
insulators and Weyl semi - metals (WSM), attracted much interests due to
their rich physics and promising prospects for application in electronic and
spinotronic devices. The band structure in the so called type I WSM like
graphene\cite{Katsnelson}, is characterized by appearance linear dispersion
relation (cones around several Dirac points) due to the band inversion. This
is qualitatively distinct from conventional metals, semi - metals or
semiconductors, in which bands are typically parabolic. In type-II WSM \cite%
{Soluyanov}, the cones have such a strong tilt, $\kappa $, so that they
exhibit a nearly flat band and the Fermi surface "encircles" the Brillouin
zone, Fig.1b, Fig.1c. It is topologically distinct from conventional
"pockets", see Fig.1a. This in turn leads to exotic electronic properties
different from both the those in both the conventional and in the type I
WSM. Examples include the collapse of the Landau level spectrum in
magnetoresistance \cite{Yu}, and novel quantum oscillations \cite{Brien}.

The type II topology of the Fermi surface was achieved in particular in
transition metal dichalcogenides \cite{Wang}. Very recently$MoTe_{2}$ layers
intercalated by ionic liquid cations were studied\cite{Zhang22}. The tilt
value was estimated to as high as $\kappa =1.3$ that places it firmly within
the type II WSM class. The measurements included the Hall effect and the
resistivity at low temperatures demonstrating appearance of
superconductivity. They discovered two intriguing facts that are currently
under discussion. First changing the gate voltage (chemical potential)
surprisingly has no impact on critical temperature, $T_{c}$, in wide range
of density of the electron gas. Second $T_{c}$ turned out to be very
sensitive to the inter - layer distance $d$: it increases from $10.5A$ to $%
11.7A$, while the critical temperature jumps from $4.2K$ to $7K$. In the
present paper we propose a theoretical explanation of these observations
based on appropriate generalization of the conventional superconductivity
theory applied to these materials.

Although early on unconventional mechanisms of superconductivity in WSM have
been considered, accumulated experimental evidence points towards the
conventional phonon mediated one \cite{DasSarma,FuBerg,frontiers}. In \ the
previous paper\cite{Rosenstein17} and a related work\cite{Zyuzin} a
continuum theory of conventional superconductivity in WSM was developed.
Magnetic response in the superconducting state was calculated\cite{Zyuzin}%
\cite{Rosenstein18}. The model was too "mesoscopic" to describe the type II
phase since the \textit{global} topology of the Brillouin zone was beyond
the scope of the continuum approach. Therefore we go beyond the continuum
model in the present paper by modeling a type II layered WSM using a tight
binding approach. The in-plane electron liquid model is similarl to that of
graphene oxide\cite{Jian12} and other 2D WSM. It possesses a chiral symmetry
between two Brave sublattices for all values of the tilt parameter $\kappa $%
, but lacks hexagonal symmetry. The second necessary additional feature is
inclusion of Coulomb repulsion.

It turns out that the screened Coulomb repulsion significantly opposes the
phonon mediated pairing. Consequently a detailed RPA theory of screening in
a layered material\cite{Elliasson} is applied. We calculate the
superconducting critical temperature taking into consideration the
modification of the Coulomb interaction due to the dielectric constant of
intercalator material and the inter-layered spacing $d$. The Gorkov
equations for the two sublattices system are solved without resorting to the
mesoscopic approach. Moreover since screening of Coulomb repulsion plays a
much more profound role in quasi 2D materials the pseudo-potential
simplification developed by McMillan\cite{McMillan} is not valid.

Rest of the paper is organized as follows. In Section II the microscopic
model of the layered WSM is described. The RPA calculation of both the
intra- and inter - layer screening is presented. In Section III the Gorkov
equations for the optical phonon mediated intra- layer pairing for a
multiband system including the Coulomb repulsion is derived and solved
numerically. In Section IV the phonon theory of pairing including the
Coulomb repulsion for a layered material is applied to recent extensive
experiments on $MoTe_{2}$. The effect of intercalation and density on
superconductivity is studied. This explains the both remarkable features of $%
T_{c}$ observed\cite{Zhang22} in $MoTe_{2}$. The last Section contains
conclusions and discussion.

\begin{figure}[h]
\centering \includegraphics[width=12cm]{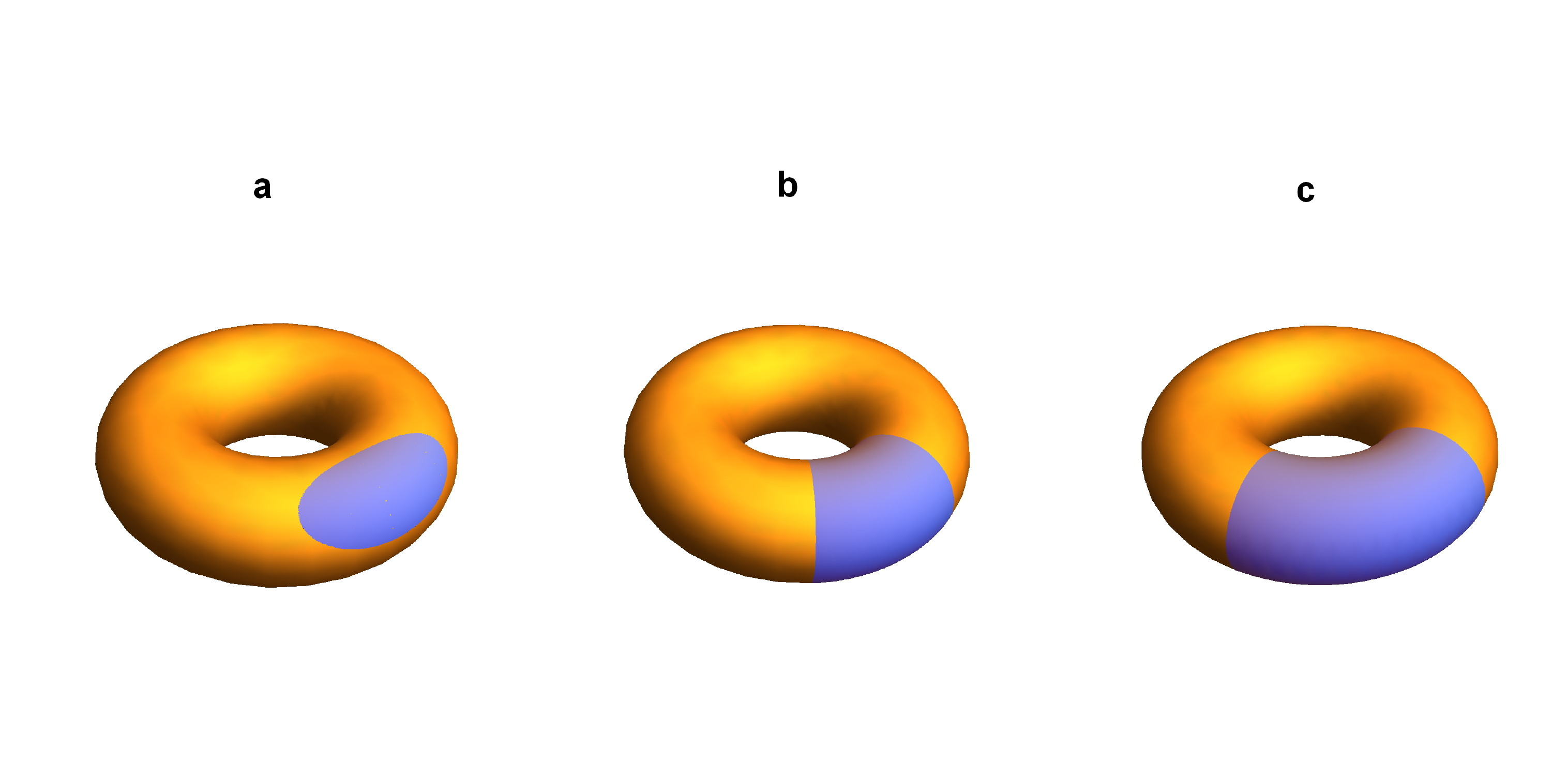}
\caption{Two distict topologies of the Fermi surface in 2D. Topology of the
2D Brillouin zone is that of the surface of 3D torroid. On the left the
``conventional'' type I pocket is shown. In the ceter and on the right the
type II topology is shown schematically. The filled states are in blue and
envelop the torus. Despite the large difference in density of the two the
Fermi surface properties like density of states are the same.}
\end{figure}

\section{A "generic" lattice model of layered Weyl semi-metals}

\subsection{Intra- layer hopping}

A great variety of tight binding models were used to describe Weyl (Dirac)
semimetals in 2D. Historically the first was graphene (type I, $\kappa =0$)
, in which electrons hope between the neighboring cites of the honeycomb
lattice. We restrict the discussion to systems with the minimal two cones of
opposite chirality and negligible spin orbit coupling. The two Dirac cones
appear in graphene at $K$ and $K^{\prime }$ crystallographic points in BZ.
Upon modification (more complicated molecules like graphene oxide, stress,
intercalation) the hexagonal symmetry is lost, however a discrete chiral
symmetry between two sublattices, denoted by $I=A,B$, ensures the WSM. The
tilted type I and even type II (for which typically $\kappa >1$) crystals
can be described by the same Hamiltonian with the tilt term added. This 2D
model is extended to a layered system with inter - layer distance $d$.
Physically the 2D WSM layers are separated by a dielectric material with
inter - layer hopping neglected, so that they are coupled
electromagnetically only\cite{Elliasson}.

The lateral atomic coordinates are still considered on the honeycomb lattice
are $\mathbf{r}_{\mathbf{n}}=n_{1}\mathbf{a}_{1}+n_{2}\mathbf{a}_{2}$, where
lattice vectors are: 
\begin{equation}
\mathbf{a}_{1}=\frac{a}{2}\left( 1,\sqrt{3}\right) ;\text{ }\mathbf{a}_{2}=%
\frac{a}{2}\left( 1,-\sqrt{3}\right) \text{,}  \label{unit cell}
\end{equation}%
despite the fact that hopping energies are different for jumps between
nearest neighbors. Each site has three neighbors separated by $\mathbf{%
\delta }_{1}=\frac{1}{3}\left( \mathbf{a}_{1}-\mathbf{a}_{2}\right) ,\mathbf{%
\delta }_{2}=-\frac{1}{3}\left( 2\mathbf{a}_{1}+\mathbf{a}_{2}\right) $ and $%
\mathbf{\delta }_{3}=\frac{1}{3}\left( \mathbf{a}_{1}+2\mathbf{a}_{2}\right) 
$, in different directions. The length of the lattice vectors $a$ will be
taken as the length unit and we set $\hbar =1$. The hopping Hamiltonian
including the tilt term is\cite{Goerbig,Jian12}:

\begin{equation}
K=\frac{\sqrt{3}}{4}\sum \nolimits_{\mathbf{n}l}\left \{ \gamma \left( \psi
_{\mathbf{n}l}^{sA\dagger }\psi _{\mathbf{r}_{\mathbf{n}}+\mathbf{\delta }%
_{1},l}^{sB}+\psi _{\mathbf{n}l}^{sA\dagger }\psi _{\mathbf{r}_{\mathbf{n}}+%
\mathbf{\delta }_{2},l}^{sB}+t\psi _{\mathbf{n}l}^{sA\dagger }\psi _{\mathbf{%
r}_{\mathbf{n}}+\mathbf{\delta }_{3},l}^{sB}\right) +\mathrm{h.c.}-\kappa
\psi _{\mathbf{n}l}^{sI\dagger }\psi _{\mathbf{r}_{\mathbf{n}}+\mathbf{a}%
_{1},l}^{sI}-\mu n_{\mathbf{n},l}\right \} \text{.}  \label{Energy}
\end{equation}%
Here\ an integer $l$ labels the layers. Operator $\psi _{\mathbf{n}%
l}^{sA\dagger }$ is the creation operators with spin $s=\uparrow ,\downarrow 
$, while the density operator is defined as $n_{\mathbf{n}l}=\psi _{\mathbf{n%
}l}^{sI\dagger }\psi _{\mathbf{n}l}^{sI}$. The chemical potential is $\mu $,
while $\gamma $ is the hopping energy for two neighbors at $\mathbf{\delta }%
_{1},\mathbf{\delta }_{2}$ . Since the the system does not possesses
hexagonal symmetry (only the chiral one), the third jump has the different
hopping\cite{Jian12} $t\gamma $. Dimensionless parameter $\kappa $
determines the tilt of the Dirac cones along the $\mathbf{a}_{1}$direction%
\cite{Goerbig}. In the 2D Fourier space, $\psi _{\mathbf{n}%
l}^{sA}=N_{s}^{-2}\sum \nolimits_{\mathbf{k}}\psi _{\mathbf{k}l}^{sA}e^{-i%
\mathbf{k\cdot r}_{n}}$, one obtains for Hamiltonian \bigskip (for finite
discrete reciprocal lattice $N_{s}\times N_{s}$):

\begin{equation}
K=N_{s}^{-2}\sum \nolimits_{\mathbf{k}l}\psi _{\mathbf{k}l}^{s\dagger }M_{%
\mathbf{k}}\psi _{\mathbf{k}l}^{s}.
\end{equation}%
Here $\mathbf{k}=\frac{k_{1}}{N_{s}}\mathbf{b}_{1}+\frac{k_{2}}{N_{s}}%
\mathbf{b}_{2}$ (reciprocal lattice vectors are given in Appendix A) and
matrix $M_{\mathbf{k}}=d_{x}\sigma _{x}+d_{y}\sigma _{y}+d_{0}I$ in terms of
Pauli matrices has components: 
\begin{eqnarray}
d_{x} &=&\frac{2t}{\sqrt{3}}\cos \left[ \frac{2\pi }{3N_{s}}\left(
k_{1}-k_{2}\right) \right] +\frac{4}{\sqrt{3}}\cos \left[ \frac{\pi }{N_{s}}%
\left( k_{1}+k_{2}\right) \right] \cos \left[ -\frac{\pi }{3N_{s}}\left(
k_{1}-k_{2}\right) \right] ;  \label{d_def} \\
d_{y} &=&-\frac{2t}{\sqrt{3}}\sin \left[ \frac{2\pi }{3N_{s}}\left(
k_{1}-k_{2}\right) \right] +\frac{4}{\sqrt{3}}\cos \left[ \frac{\pi }{N_{s}}%
\left( k_{1}+k_{2}\right) \right] \sin \left[ \frac{\pi }{3N_{s}}\left(
k_{1}-k_{2}\right) \right] ;  \notag \\
d_{0} &=&\frac{2}{\sqrt{3}}\left \{ -\kappa \cos \left[ \frac{2\pi }{N_{s}}%
k_{1}\right] -\mu \right \} \text{.}  \notag
\end{eqnarray}

Using $\gamma $ as our energy unit from now on, the free electrons part of
the Matsubara action for Grassmanian fields $\psi _{\mathbf{k}ln}^{\ast sI}$
is:

\begin{equation}
S^{e}=\frac{1}{T}\sum \nolimits_{\mathbf{k}ln}\psi _{\mathbf{k}ln}^{\ast
sI}\left \{ \left( -i\omega _{n}+d_{\mathbf{k}}^{0}\right) \delta
^{IJ}+\sigma _{i}^{IJ}d_{\mathbf{k}}^{i}\right \} \psi _{\mathbf{k}ln}^{sJ}%
\text{.}  \label{Action_e}
\end{equation}%
where $\omega _{n}=\pi T\left( 2n+1\right) $ is the Matsubara frequency. The
Green Function of free electrons has the matrix form

\begin{equation}
g_{\mathbf{k}n}=\left \{ \left( -i\omega _{n}+d_{\mathbf{k}}^{0}\right)
I+\sigma _{i}d_{\mathbf{k}}^{i}\right \} ^{-1}=\frac{\left( -i\omega _{n}+d_{%
\mathbf{k}}^{0}\right) I-\sigma _{i}d_{\mathbf{k}}^{i}}{\left( i\omega
_{n}-d_{\mathbf{k}}^{0}\right) ^{2}-d_{\mathbf{k}}^{x2}-d_{\mathbf{k}}^{y2}}%
\text{.}  \label{gdef}
\end{equation}%
Now we turn to the spectrum of this model.

\begin{figure}[h]
\centering \includegraphics[width=8cm]{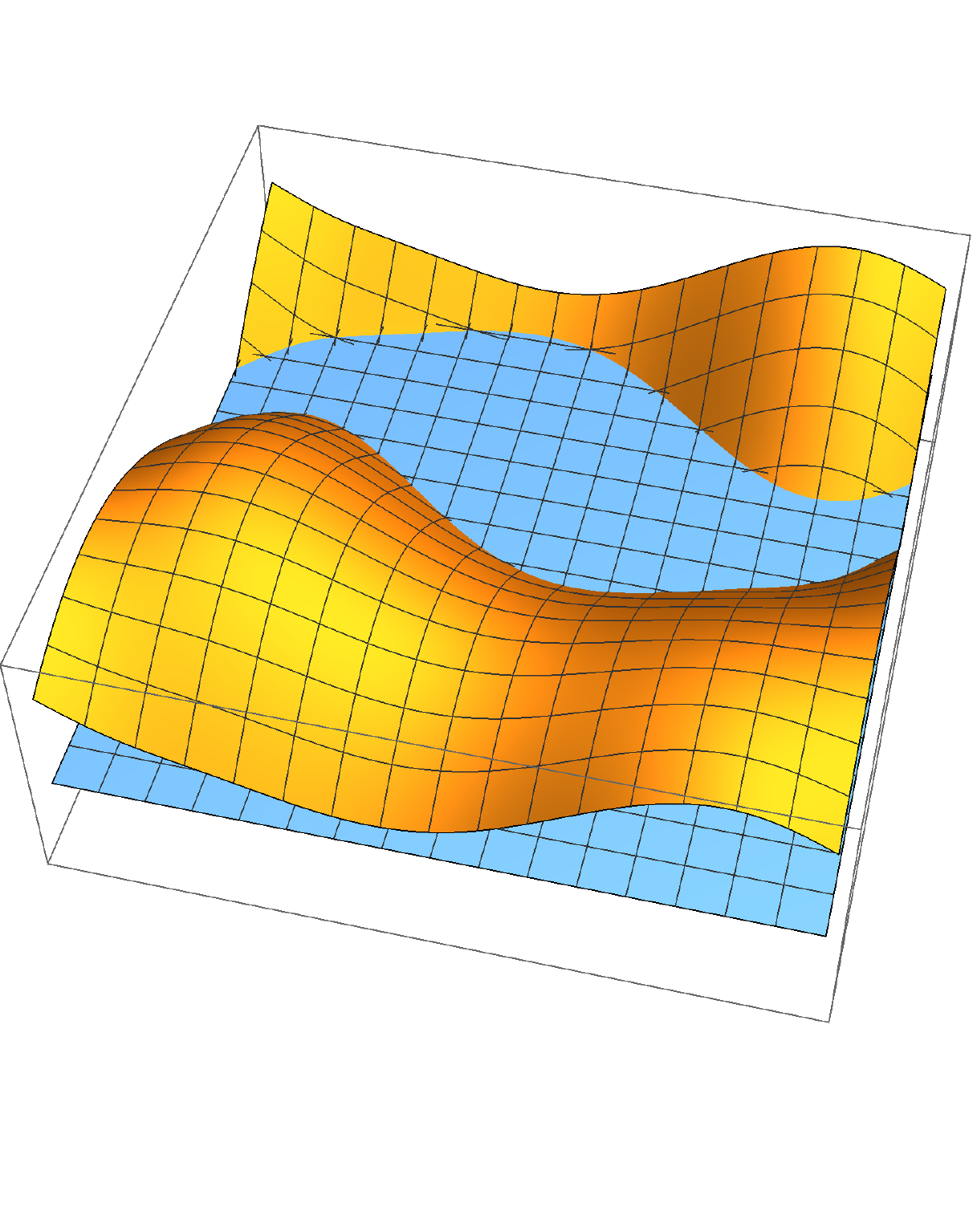}
\caption{The topological phase diagram of the Weyl semimetal at large tilt
parameter ($\protect\kappa =1.3$). Chemical potential (in units of $\protect%
\gamma =500$ meV) is marked on each contour. The electron type I topology at
low values of $\protect\mu $ undergoes transition to the type II at $\protect%
\mu =\protect\mu _{c}^{1}=0.8$ meV. At yet larger $\protect\mu >\protect\mu %
_{c}^{2}=1.35$. the Fermi surface becomes again type I. This time the
excitations are hole rather than electrons.}
\end{figure}

\begin{figure}[h]
\centering \includegraphics[width=8cm]{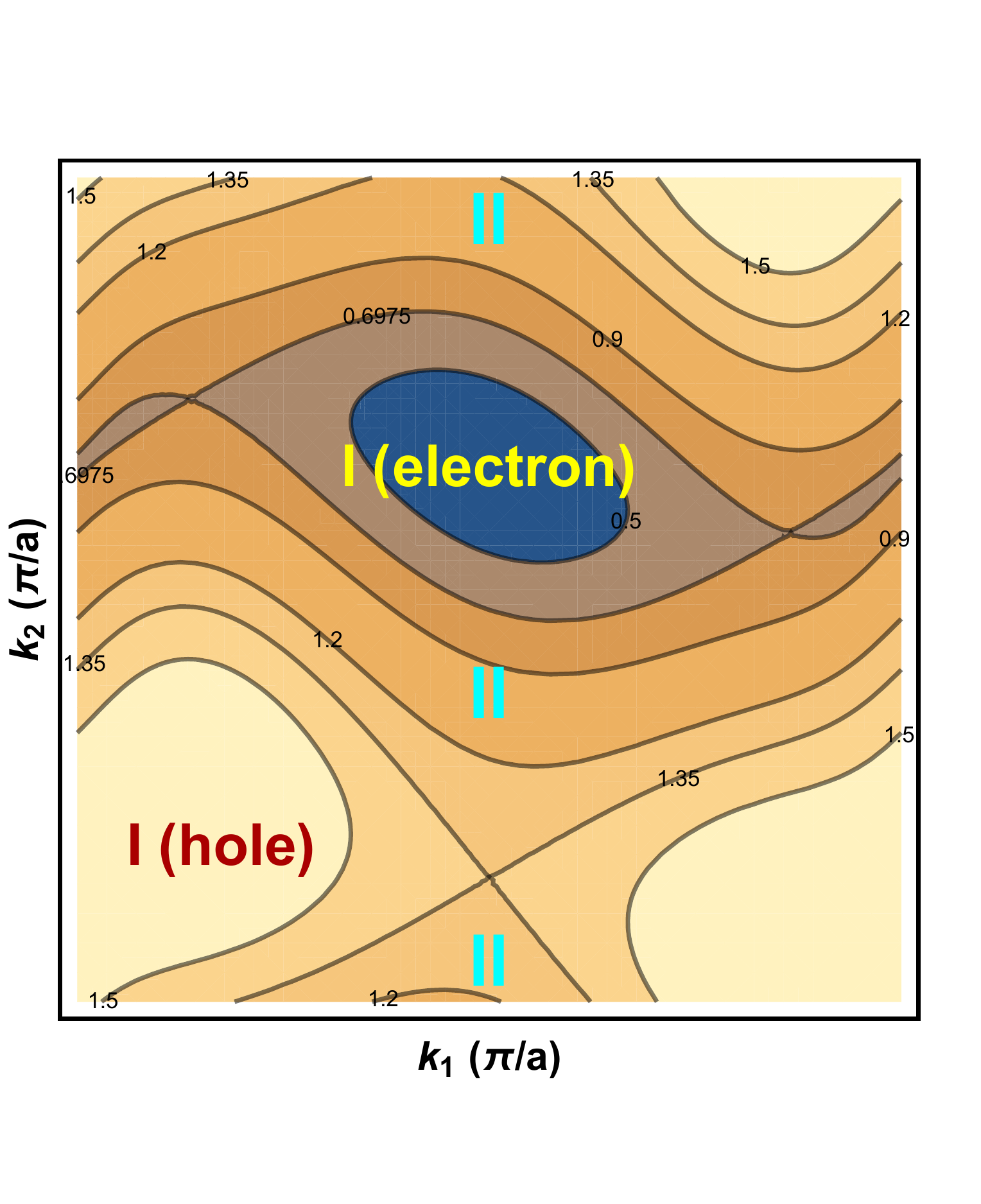}
\caption{Dispersion relation of WSM with $\protect\kappa =1.3$. The blue
plane corresponds to chemical potentia l$\protect\mu =0.8$ eV so that the
Fermi surface has the type II topology. }
\end{figure}

\subsection{The range of the topological type II phase at large $\protect%
\kappa $}

The spectrum of Hamiltonian of Eqs.(\ref{d_def}) consists of two branches.
The upper branch for $\mu =0.9eV$ is given in Fig. 2. The lower branch for a
reasonable choice of parameters appropriate to $MoTe_{2}$ is significantly
below the Fermi surface and is not plotted. Blue regions represent the
filled electron states. One observes a "river" from one boundary to the
other of the Brillouin zone (in coordinates $k_{1}$ and $k_{2}$, in terms of
the original $k_{x},k_{y}$ it is a rhomb) characteristic to type II Fermi
surface. Topologically this is akin to Fig.1b.

In Fig. 3 the Fermi surfaces in a wide range of densities $n=7.\times
10^{13}-4.5\times 10^{14}cm^{-2}$ are given. Topologically they separate
into three phases. At chemical potentials below $\mu _{c}^{1}=0.796$ $eV$,
corresponding to densities $n<n_{c}^{1}=8.\times 10^{13}$ $cm^{-2}$ , the
Fermi surface consists of one compact electron pocket similar to Fig.1a, so
that the electronic matter is of the ("customary") topological type I. The
density is determined from the (nearly linear) relation between the chemical
potential and density given in Fig. 4 (blue line, scale on the right). In
the range $\mu _{c}^{1}<\mu <\mu _{c}^{2}=1.35eV$ the Fermi surface consists
of two banks of a "river" (blue color represents filled electron states) in
Fig.2 and can be viewed topologically as in Fig.1b and Fig1c. The second
critical density is $n_{c}^{2}=3.6\times 10^{14}$ $cm^{-2}$. In this range
the shape of both pieces of the Fermi surface largely does not depend on the
density that is proportional to the area of the blue part of the surface.

To make this purely topological observation quantitative, we present in Fig.
4 (green line, scale on the left) the density of states (DOS) as a function
of chemical potential. One observes that it nearly constant away from the
two topological I to II transitions where it peaks.

\begin{figure}[h]
\centering \includegraphics[width=8cm]{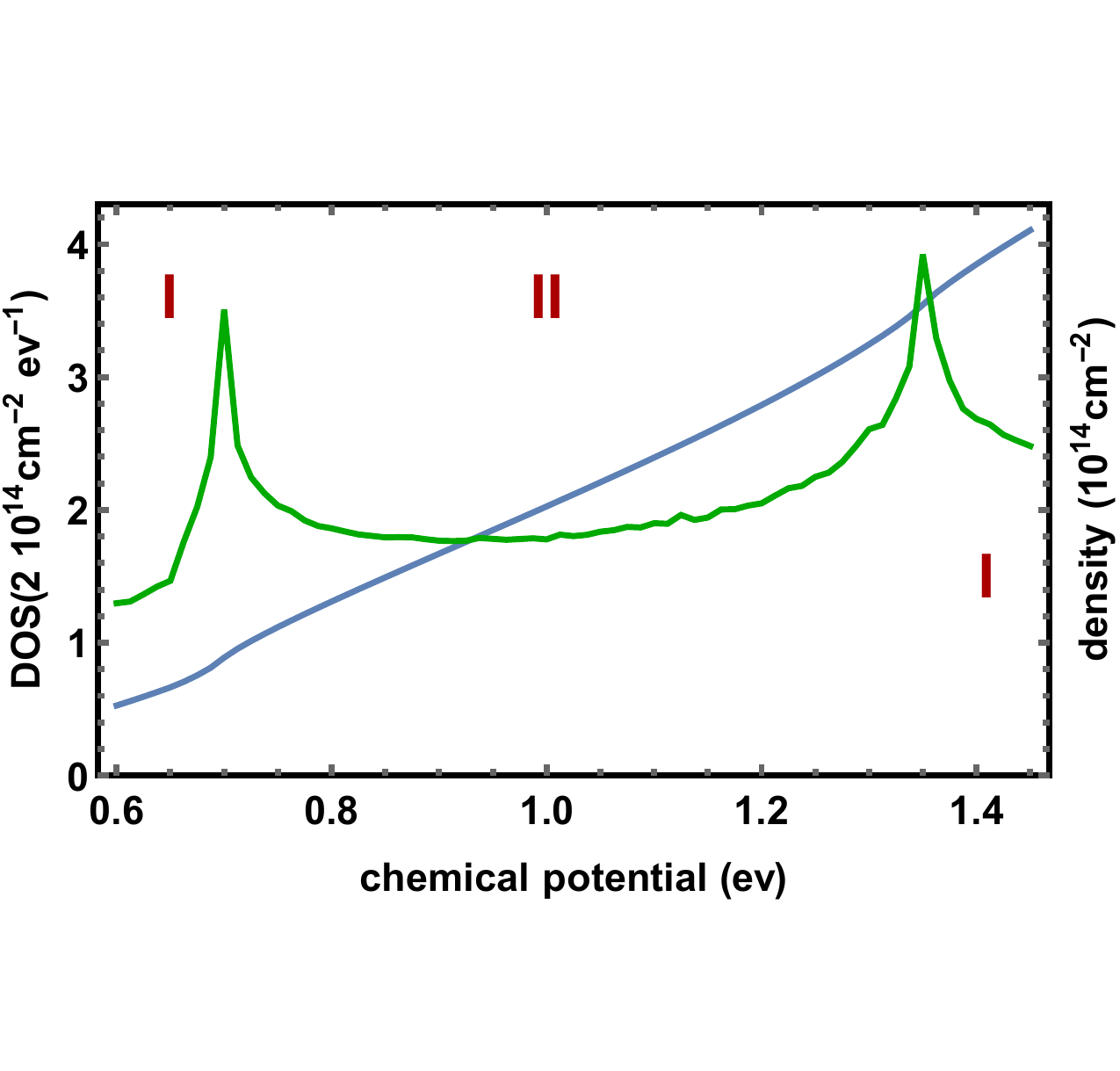}
\caption{Electron density and DOS as function of the chemical potential $%
\protect\mu $.of WSM with $\protect\kappa =1.3$. DOS has cusps at both I to
II transitions. Between the transitions it is nearly constant in the range
of densities from $1.1\times 10^{14}/cm^{2}$ to $4.\times 10^{14}/cm^{2}$. }
\end{figure}

\subsection{Coulomb repulsion}

The electron-electron repulsion in the layered WSM can be presented in the
form,

\begin{equation}
V=\frac{e^{2}}{2}\sum \nolimits_{\mathbf{n}l\mathbf{n}^{\prime }l^{\prime
}}n_{\mathbf{n}l}v_{\mathbf{n-n}^{\prime },l-l^{\prime }}^{C}n_{\mathbf{n}%
^{\prime }l^{\prime }}=\frac{e^{2}}{2N_{s}^{2}}\sum \nolimits_{\mathbf{q}%
ll^{\prime }}n_{\mathbf{q}l}n_{-\mathbf{q}l^{\prime }}v_{\mathbf{q,}%
l-l^{\prime }}^{C}\text{,}  \label{Coulomb}
\end{equation}%
where $v_{\mathbf{n-n}^{\prime },l-l^{\prime }}^{C}$ is the "bare" Coulomb
interaction between electrons with Fourier transform $v_{\mathbf{q}%
,l-l^{\prime }}^{C}=v_{\mathbf{q}}^{2D}e^{-dq\left \vert l-l^{\prime
}\right
\vert }$, $v_{\mathbf{q}}^{2D}=2\pi e^{2}/q\epsilon $. Here $%
\epsilon $ is the dielectric constant of the intercalator material

The long range Coulomb interaction is effectively taken into account using
the RPA approximation.

\section{Screening in layered WSM.}

The screening in the layered system can be conveniently partitioned into the
screening within each layer described by the polarization function $\Pi _{%
\mathbf{q}n}$ and electrostatic coupling to carriers in other layers. We
start with the former.

\subsection{Polarization function of the electron gas in Layered WSM}

In a simple Fermi theory of the electron gas in normal state with Coulomb
interaction between the electrons in RPA approximation the Matsubara
polarization is calculated as a simple \textit{minus} "fish" diagram \cite%
{Elliasson} in the form:

\begin{equation}
\Pi _{\mathbf{q}n}=-\left( -2T\text{Tr}\sum \nolimits_{\mathbf{p}m}g_{%
\mathbf{p}m}g_{\mathbf{p+q},m+n}\right) \text{.}  \label{Polarization}
\end{equation}%
Using the GF of Eq.(\ref{gdef}), one obtain:

\begin{equation}
\Pi _{\mathbf{q}n}=4T\sum \nolimits_{\mathbf{p}m}\frac{\left( i\omega
_{m}+A\right) \left( i\omega _{m}+B\right) +C}{\left[ \left( i\omega
_{m}+A\right) ^{2}-\alpha ^{2}\right] \left[ \left( i\omega _{m}+B\right)
^{2}-\beta ^{2}\right] }\text{,}  \label{Polarization omega}
\end{equation}%
where

\begin{eqnarray}
A &=&-d_{\mathbf{p}}^{0};B=i\omega _{n}-d_{\mathbf{p+q}}^{0};\text{ \ \ }%
C=d_{\mathbf{p}}^{x}d_{\mathbf{p+q}}^{x}+d_{\mathbf{p}}^{y}d_{\mathbf{p+q}%
}^{y};  \label{parameters} \\
\alpha ^{2} &=&d_{\mathbf{p}}^{x2}+d_{\mathbf{p}}^{y2};\text{ \ \ }\beta
^{2}=d_{\mathbf{p+q}}^{x2}+d_{\mathbf{p+q}}^{y2}\text{.}  \notag
\end{eqnarray}%
Performing summation over $m$, one obtains:

\begin{equation}
\Pi _{\mathbf{q}n}=-\sum \nolimits_{\mathbf{p}}\left \{ 
\begin{array}{c}
\frac{\alpha ^{2}-\alpha (A-B)+C}{\alpha \left[ \left( A-B-\alpha \right)
^{2}-\beta ^{2}\right] }\tanh \frac{\alpha -A}{2T}+\frac{a^{2}+\alpha (A-B)+C%
}{\alpha \left[ \left( A-B+\alpha \right) ^{2}-\beta ^{2}\right] }\tanh 
\frac{\alpha +A}{2T} \\ 
+\frac{\beta ^{2}+\beta \left( A-B\right) +C}{\beta \left[ \left( A-B+\beta
\right) ^{2}-\alpha ^{2}\right] }\tanh \frac{\beta -B}{2T}+\frac{\beta
^{2}-\beta \left( A-B\right) +C}{\beta \left[ \left( A-B-\beta \right)
^{2}-\alpha ^{2}\right] }\tanh \frac{\beta +B}{2T}%
\end{array}%
\right \} \text{.}  \label{Polar}
\end{equation}%
Now we turn to screening due to other layers.

\subsection{Screening in a layered system}

Coulomb repulsion between electrons in different layers $l$ and $l^{\prime }$
within the RPA approximation is determined by the following integral
equation:

\begin{equation}
V_{\mathbf{q,}l-l^{\prime }\mathbf{,}n}^{RPA}=v_{\mathbf{q},l-l^{\prime
}}^{C}+\Pi _{\mathbf{q}n}\sum \nolimits_{l^{\prime \prime }}v_{\mathbf{q}%
,l-l^{\prime \prime }}^{C}V_{\mathbf{q},l^{\prime \prime }-l^{\prime }%
\mathbf{,}n}^{RPA}\text{.}  \label{Series}
\end{equation}%
The polarization function $\Pi _{\mathbf{q}n}$ in 2D was calculated in the
previous subsection. This set of equations is decoupled by the Fourier
transform in the $z$ direction,

\begin{equation}
V_{\mathbf{q,}q_{z},n}^{RPA}=\frac{v_{\mathbf{q},q_{z}}^{C}}{1-\Pi _{\mathbf{%
q}n}v_{\mathbf{q},q_{z}}^{C}}\text{ , \ }  \label{RPA}
\end{equation}%
where%
\begin{equation}
v_{\mathbf{q},q_{z}}^{C}=\sum \nolimits_{l}v_{\mathbf{q}}^{2D}e^{iq_{z}l-qd%
\left \vert l\right \vert }=v_{\mathbf{q}}^{2D}\frac{\sinh \left[ qd\right] 
}{\cosh \left[ qd\right] -\cos \left[ dq_{z}\right] }\text{.}  \label{vCdef}
\end{equation}%
The screened interaction in a single layer therefore is is given by the
inverse Fourier transform \cite{Elliasson}:

\begin{equation}
V_{\mathbf{q,}l-l^{\prime },n}^{RPA}=\frac{d}{2\pi }\int_{q_{z}=-\pi
/d}^{\pi /d}e^{iq_{z}d\left( l-l^{\prime }\right) }\frac{v_{\mathbf{q}%
,q_{z}}^{C}}{1-\Pi _{\mathbf{q}n}v_{\mathbf{q},q_{z}}^{C}}\text{.}
\label{Screening}
\end{equation}%
Considering screened Coulomb potential at the same layer $l=l^{\prime },\ $%
the integration gives,

\begin{equation}
V_{\mathbf{q}n}^{RPA}=\frac{v_{\mathbf{q}}^{2D}\sinh \left[ qd\right] }{%
\sqrt{b_{\mathbf{q}n}^{2}-1}}\text{,}  \label{inlayer repulsion}
\end{equation}%
where $b_{\mathbf{q}n}=\cosh \left[ dq\right] -v_{\mathbf{q}}^{2D}\Pi _{%
\mathbf{q}n}\sinh \left[ dq\right] $. This formula is reliable only away
from plasmon region $b_{\mathbf{q}n}>1$. It turns out that to properly
describe superconductivity, one can simplify the calculation at low
temperature by considering the static limit $\Pi _{\mathbf{q}n}\simeq \Pi _{%
\mathbf{q}0}$. Consequently the potential becomes static: $V_{\mathbf{q}%
}^{RPA}\equiv V_{\mathbf{q},n=0}^{RPA}$.

\section{Superconductivity}

Superconductivity in WSM is caused by a conventional phonon pairing. The
leading mode is an optical phonon mode assumed to be dispersionless. with
energy $\Omega $. The effective electron-electron attraction due to the
electron - phonon attraction opposed by Coulomb repulsion (pseudo -
potential) mechanism creates pairing below $T_{c}$. Further we assume the
singlet $s$-channel electron-phonon interaction and neglect the inter-layers
electrons pairing. \ In order to describe superconductivity, one should
"integrate out" the phonon and the spin fluctuations degrees of freedom to
calculate the effective electron - electron interaction. We start with the
phonons. The Matsubara action for effective electron-electron interaction
via in-plane phonons and direct Coulomb repulsion calculated in the previous
Section. It important to note that unlike in metal superconductors where a
simplified pseudo - potential approach due to McMillan and other \cite%
{McMillan}, in 2D and layered WSM, one have to resort to a more microscopic
approach.

\subsection{Effective attraction due to phonon exchange opposed by the
effective Coulomb repulsion}

The free and the interaction parts of the effective electron action
("integrating phonons"+RPA Coulomb interaction) in the quasi - momentum -
Matzubara frequency representation, $S=S^{e}+S^{int}$,

\begin{equation}
S^{int}=\frac{1}{2T}\sum \nolimits_{\mathbf{q}ll^{\prime }mm^{\prime }}n_{%
\mathbf{q}ln}\left( \delta _{ll^{\prime }}V_{\mathbf{q,}m-m^{\prime
}}^{ph}+V_{\mathbf{q},l-l^{\prime }}^{RPA}\right) n_{-\mathbf{q},-l^{\prime
},-n^{\prime }}\text{.}  \label{Interaction action}
\end{equation}%
Here $n_{\mathbf{q}ln}=\sum \nolimits_{\mathbf{p}}\psi _{\mathbf{p}ln}^{\ast
sI}\psi _{\mathbf{q-p,}l,n}^{sI}$ the Fourier transform of the electron
density and $S^{e}$ was defined in Eq.(\ref{Action_e}). The effective
electron - electron coupling due to phonons is:

\begin{equation}
\text{ \ }V_{\mathbf{q}m}^{ph}=-\left( \frac{\sqrt{3}}{2}\right) ^{2}\frac{%
g^{2}\Omega }{\omega _{m}^{b2}+\Omega ^{2}}\text{,}
\label{electron phonon plus Coulomb}
\end{equation}%
where the bosonic frequencies are $\omega _{m}^{b}=2\pi mT$.

\subsection{Gorkov Green's functions and the s-wave gap equations}

Normal and anomalous (Matsubara) intra - layer Gorkov Green's functions are
defined by expectation value of the fields, $\left \langle \psi _{\mathbf{k}%
nl}^{Is}\psi _{\mathbf{k}nl}^{\ast s^{\prime }J}\right \rangle =\delta
^{ss^{\prime }}G_{\mathbf{k}n}^{IJ}$ and $\left \langle \psi _{\mathbf{k}%
nl}^{Is}\psi _{-\mathbf{k,-}n,l}^{Js^{\prime }}\right \rangle =\varepsilon
^{ss^{\prime }}F_{\mathbf{k}n}^{IJ}$, while the gap function is%
\begin{equation}
\Delta _{\mathbf{q}n}^{IJ}=\sum \nolimits_{\mathbf{p}m}V_{\mathbf{q-p,}%
n-m}F_{\mathbf{p}m}^{IJ},  \label{deltadef}
\end{equation}%
where $V_{\mathbf{q}n}=V_{\mathbf{q}n}^{ph}+V_{\mathbf{q}n}^{RPA}$ is a
sublattice scalar. The gap equations in the sublattice matrix form are
derived from Gorkov equations in Appendix B:

\begin{equation}
\Delta _{\mathbf{q}n}=-\sum \nolimits_{\mathbf{p}m}V_{\mathbf{q-p,}n-m}g_{%
\mathbf{p}m}\left \{ I+\Delta _{\mathbf{p}m}g_{-\mathbf{p},-m}^{t}\Delta _{-%
\mathbf{p},-m}^{\ast }g_{\mathbf{p}m}\right \} ^{-1}\Delta _{\mathbf{p}m}g_{-%
\mathbf{p},-m}^{t}\text{.}  \label{Gap}
\end{equation}

In numerical simulation the gap equation was solved iteratively. Relatively
large space cutoff $N_{s}=256$ is required. The frequency cutoff $N_{t}=128$
was required due to low temperatures approached. Typically $15-25$
iterations were required. The parameters used were $\Omega =16meV$. The
electron - phonon coupling $g=20meV$. Now we turn to results concentrating
on two puzzling experimental results of ref.\cite{Zhang22}.

\subsection{Independence of $T_{c}$ on density in topological type II phase}

In Fig.5 the critical temperature for various values of density are
plotted.\bigskip \ The blue points are for dielectric constant\cite{Zhang22}%
, $\varepsilon =16$, describing the intercalated imidazole cations $%
[C_{2}MIm]$ \cite{dc}. The inter - layer distance was kept at $d=10.5A$.

\begin{figure}[h]
\centering \includegraphics[width=8cm]{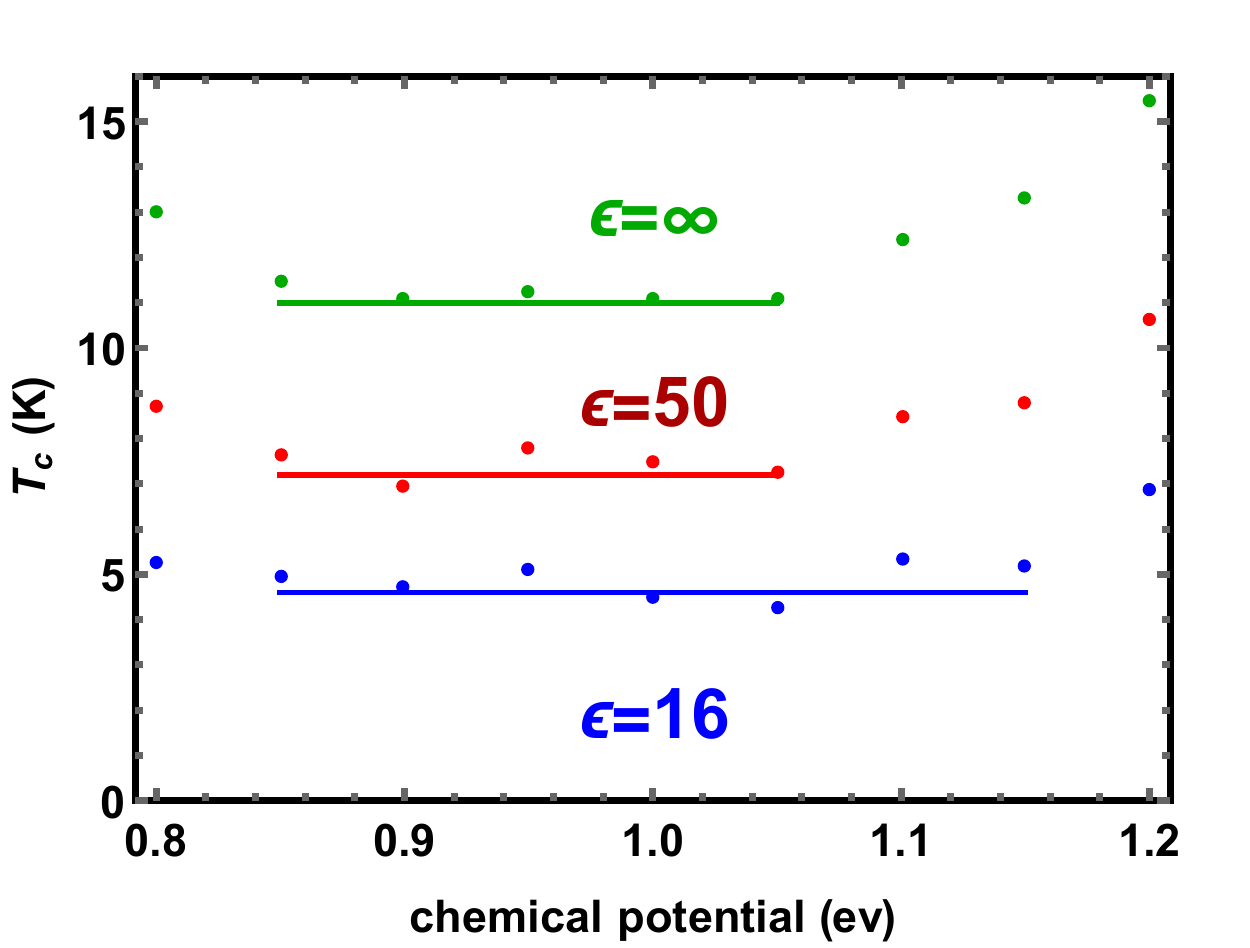}
\caption{Critical temperature of transition to superconducting state in type
II layered WSM is shown as function of chemical potential (can be translated
into carrier density via Fig.4). Three values of dielectric constant of the
intercalant for fixed interlayer distance are shown. Parameters of the
electron gas are the same as in previous figures.}
\end{figure}
The significance and generatiozation of the observation are discussed below.

\subsection{Increase of $T_{c}$ with \ dielectric constant of intercalator
materials}

\ The main idea of the paper is that the difference in $T_{c}$ between
different intercalators is attributed not to small variations in the inter -
layer spacing $d$, but rather to large differences in the dielectric
constant of the intercalating materials due to its effect on the screening.
In experiment of ref.\cite{Zhang22} the imidazole cations $[C_{2}MIm]^{+}$
(1- ethyl - 3 - methyl - imidazolium) are short molecules\cite{dc} have $%
\epsilon =16$, while $[C_{6}MIm]^{+}$ (1- hexy l - 3 - methy l -
imidazolium) are long molecules\cite{dc2} with a larger value $\epsilon
\simeq 50$. The inter - layer distance $d$ is slightly dependent
intercalators changing from $10.5\mathring{A}$ to $11.7\mathring{A}$ . The
blue points in Fig 5 describe a material with dielectric constant $%
\varepsilon =16$ should. This is contrasted\cite{dc2} with the $\varepsilon
=50$ material, see the red point. Neglecting the Coulomb repulsion, see the
green points, critical temperature (a much simpler calculation of $T_{c}$ in
this case similar to that in ref.\cite{Rosenstein17} is needed in this case)
becomes yet higher. This demonstrate the importance of the Coulomb repulsion
in a quasi 2D system. Superconductivity is weaker for monolayer on substrate
since both air and substrate have smaller dielectric constants and hence
weaker screen the Coulomb repulsion.

\section{Discussion and conclusion}

To summarize we have developed a theory of superconductivity in layered type
II Weyl semi-metals that properly takes into account the Coulomb repulsion.
The generalization goes beyond the simplistic pseudo - potential approach
due to McMillan\cite{McMillan} and others and depends essentially on the
intercalating material. The theory allows to explain the two puzzling
phenomena observed recently in layered intercalated $MoTe_{2}$ WSW compound 
\cite{Zhang22}

The first experimental observation is that the gate voltage (changes in the
chemical potential or equivalently in density) has no impact on critical
temperature $T_{c}$. For the 3D density range $8.\times
10^{20}cm^{-3}-3.6\times 10^{21}cm^{-3}$ the temperature changes within $5\%$%
. For the intercalating material $[C_{2}MIm]^{+}$ with inter - layer
distance $d=10.5A$ the 2D density range translates into $8.4\times
10^{13}cm^{-2}-3.8\times 10^{14}cm^{-2}$ wth slightly larger spacings $%
d=11.7A$ shown in Figs 2-5. This feature is explained purely topologically,
see schematic Fig.1. In the type II density range the shape of both pieces
of the Fermi surface (the blue - yellow boundaries in Fig1b and Fig1c)
largely does not depend on the density (that is proportional to the area of
the blue part of the surface) leading, see Fig. 4 to approximate
independence of the density of states (DOS) $N\left( 0\right) $ of chemical
potential $\mu $. This feature is akin to the DOS independence on $\mu $ for
a parabolic (topologically type I like in Fig.1a) band in purely 2D
materials, but has completely difficult origin.

Using the somewhat naive BCS formula%
\begin{equation}
T_{c}\simeq \Omega \text{ }e^{-N\left( 0\right) g_{eff}^{2}}\text{.}
\end{equation}%
Here $\Omega $ is the phonon frequency and $g_{eff}$ the effective electron
- phonon coupling. Assuming that both $\Omega $ and $g$ do not depend on the
density one arrives at a conclusion that in the type II topological phase
the critical temperature is density independent .

The second experimental observation\cite{Zhang22} was that $T_{c}$ is in
fact very sensitive to the intercalating material. For imidazole cations $%
[C_{2}MIm]^{+}$ the critical temperature is $T_{c}=4.2K$, while for $%
[C_{6}MIm]^{+}$ the temperature jumps to $T_{c}=6.6K$ or $6.9K$ depending on
the intercalation method. The inter - layer distance $d$ is slightly
dependent intercaltors increasing from $10.5\mathring{A}$ to $11.7\mathring{A%
}$ . Our calculation demonstrates that the difference in $T_{c}$ between
different intercalators cannot be attributed to small variations in the
inter - layer spacing $d$. On the contrary there are large differences in
the dielectric constant of the intercalating materials. While $%
[C_{2}MIm]^{+} $ have\cite{dc} a relatively small dielectric constant $%
\epsilon =16$, $[C_{6}MIm]^{+}$ is estimated\cite{dc2} in the range $%
\epsilon =40-60$. Our theory accounts the difference in $T_{c}$ due to
changes in the screening of the Coulomb potential due to the inter - layer
insulator.

\section{\protect\bigskip \textit{Acknowledgements. }}

This work was supported by NSC of R.O.C. Grants No. 101-2112-M-009-014-MY3.

\section{\protect\bigskip}

\section{Appendix A. Details of the model}

The system considered in the paper is fitted for the following values of the
hipping and the tilt parameter. The dimensionless tilt parameter was taken
from ref.\cite{Zhang22} $\kappa =1.3$. The hopping $\gamma =500$ $meV$ and $%
t=2$. The calculations were performed on the discrete reciprocal lattice $%
k_{1},k_{2}=1,...N_{s}$ with $N_{s}=256$. Reciprocal lattice basis vectors
are,%
\begin{equation}
\mathbf{b}_{1}=2\pi \left( 1,\frac{1}{\sqrt{3}}\right) ;\text{ \ \ \ }%
\mathbf{b}_{2}=2\pi \left( 1,-\frac{1}{\sqrt{3}}\right) ,  \label{SM2}
\end{equation}%
so that a convenient representation is $\mathbf{k}=\frac{k_{1}}{N_{s}}%
\mathbf{b}_{1}+\frac{k_{2}}{N_{s}}\mathbf{b}_{2}$ with 
\begin{equation}
k_{x}=\frac{2\pi }{N_{s}}\left( k_{1}+k_{2}\right) ,\text{ \ \ }k_{y}=\frac{%
2\pi }{\sqrt{3}N_{s}}\left( k_{1}-k_{2}\right) \text{.}
\end{equation}

\section{Appendix B. Derivation of the two sublattice gap equation}

\subsection{ Green's functions and the s-wave Gorkov equations}

We derive the Gorkov's equations (GE) within the functional integral approach%
\cite{NO} starting from the effective electron action for grassmanian fields 
$\psi ^{\ast X},\psi ^{Y}$.\ 
\begin{equation}
S=\frac{1}{T}\left \{ \psi ^{\ast X}\left( G_{0}^{-1}\right) ^{XY}\psi ^{Y}+%
\frac{1}{2}\psi ^{\ast Y}\psi ^{Y}V^{YX}\psi ^{\ast X}\psi ^{X}\right \} ,
\label{S Action}
\end{equation}%
where $X,Y$ denote space coordinate, sublattices (pseudospin) and spin of
the electron. Finite temperature properties of the condensate are described
at temperature $T$ by the normal and the anomalous Matsubara Greens
functions for spin singlet state.

The GE in functional form\ are:

\begin{equation}
\left \langle \psi ^{A}\psi ^{\ast B}\right \rangle \frac{\delta }{\delta
\psi ^{\ast C}}\left \langle \frac{\delta S}{\delta \psi ^{\ast B}}\right
\rangle +\left \langle \psi ^{A}\psi ^{B}\right \rangle \frac{\delta }{%
\delta \psi ^{\ast C}}\left \langle \frac{\delta S}{\delta \psi ^{B}}\right
\rangle =0;  \label{First}
\end{equation}

\begin{equation}
\left \langle \psi ^{A}\psi ^{\ast B}\right \rangle \frac{\delta }{\delta
\psi ^{C}}\left \langle \frac{\delta S}{\delta \psi ^{\ast B}}\right \rangle
+\left \langle \psi ^{A}\psi ^{B}\right \rangle \frac{\delta }{\delta \psi
^{C}}\left \langle \frac{\delta S}{\delta \psi ^{B}}\right \rangle =\delta
^{AC}\text{.}  \label{Second}
\end{equation}

Performing the calculations and using the normal and anomalous Green
functions in the form $F^{AB}=\left \langle \psi ^{A}\psi ^{B}\right \rangle
;G^{AB}=\left \langle \psi ^{A}\psi ^{\ast B}\right \rangle ,$ \bigskip one
obtains:%
\begin{equation}
F^{AX}\left \{ \left( G_{0}^{-1}\right)
^{CX}-v^{XC}G^{CX}+v^{CX}G^{XX}\right \} +G^{AX}v^{XC}F^{XC}=0.
\label{firstGE}
\end{equation}%
Skipping second and third terms in bracket in this expression and defining,
superconducting gap $\Delta ^{AB}=v^{AB}F^{AB}$, one rewrites as a matrix
products:

\begin{equation}
\left( G_{0}^{-1}\right) ^{CX}F^{XA}=G^{AX}\Delta ^{XC}\text{.}
\end{equation}%
The first GE (multiplied from left by $G_{0}$)\ is,

\begin{equation}
F^{AB}=-G^{AX}G_{0}^{BY}\Delta ^{XY}\text{,}  \label{AppB2}
\end{equation}%
while the second GE similarly is:%
\begin{equation}
G^{AB}-G^{AX}\Delta ^{XY}G_{0}^{ZY}\Delta ^{\ast ZU}G_{0}^{UB}=G_{0}^{AB}%
\text{.}  \label{AppB3}
\end{equation}

\subsection{Frequency-quasi-momentum and the spin-sublattice decomposition}

The generalized index $A$ contains the space variables (space + Matsubara
time, $a$), spin $s$ and the sublattice $I$. After performing the Fourier
series with combined quasi - momentum - frequency $\alpha $: 
\begin{eqnarray}
F_{ab}^{s_{1}s_{2}IJ} &=&\epsilon ^{s_{1}s_{2}}\sum \nolimits_{\alpha
}e^{i\alpha \left( a-b\right) }F_{\alpha }^{IJ};\text{ \ }\Delta
_{ab}^{s_{1}s_{2}IJ}=\epsilon ^{s_{1}s_{2}}\sum \nolimits_{\alpha
}e^{i\alpha \left( a-b\right) }\Delta _{\alpha }^{IJ};  \label{FourierApp} \\
G_{0ab}^{s_{1}s_{2}IJ} &=&\delta ^{s_{1}s_{2}}\sum \nolimits_{\alpha
}e^{i\alpha \left( a-b\right) }g_{\alpha }^{IJ};\text{ \ }%
V_{ab}^{s_{1}s_{2}IJ}=\sum \nolimits_{\alpha }e^{i\alpha \left( a-b\right)
}v_{\alpha }\text{.}  \notag
\end{eqnarray}

Substituting spins into Eq.(\ref{AppB2},\ref{AppB3}), one obtains in the
sublattice matrix form

\begin{eqnarray}
F_{\alpha } &=&-G_{\alpha }\Delta _{\alpha }g_{-\alpha }^{t};  \label{Gorkov}
\\
G_{\alpha } &=&G_{0\alpha }\left \{ I+\Delta _{\alpha }g_{-\alpha
}^{t}\Delta _{-\alpha }^{\ast }G_{0\alpha }\right \} ^{-1}\text{,}  \notag
\end{eqnarray}

Convoluting the first GE by $v_{\nu }$ one obtains:

\begin{equation}
\Delta _{\omega }=-\sum \nolimits_{\nu }v_{\omega -\nu }G_{\nu }\Delta _{\nu
}g_{-\nu }^{t}\text{.}
\end{equation}%
The solution of the second GE for $G$ is:

\begin{equation}
G_{\alpha }=g_{\alpha }\left \{ I+\Delta _{\alpha }g_{-\alpha }^{t}\Delta
_{-\alpha }^{\ast }g_{\alpha }\right \} ^{-1}\text{.}
\end{equation}%
Substituting into the first GE one obtain Eq.(\ref{Gap}) in the text.

\end{document}